\documentclass[twocolumn,showpacs,aps,floatfix,pra,superscriptaddress]{revtex4}
\usepackage{graphicx}
\def\la{\langle}
\def\ra{\rangle}
\def\beq{\begin{equation}}
\def\eeq{\end{equation}}
\def\beqa{\begin{eqnarray}}
\def\eeqa{\end{eqnarray}}

\begin{document}
\title{Suppression of Zeno effect for distant detectors}
\author{F. Delgado}
\email{qfbdeacf@lg.ehu.es} \affiliation{Departamento de F\'{\i}sica B\'{a}sica, Universidad de La Laguna, La Laguna, Tenerife, Spain}
\affiliation{Departamento de Qu\'\i mica-F\'\i sica, Universidad del Pa\'\i s Vasco, Apdo. 644, 48080 Bilbao, Spain}
\author{J. G. Muga}
\email{jg.muga@ehu.es} \affiliation{Departamento de Qu\'\i mica-F\'\i sica, Universidad del Pa\'\i s Vasco, Apdo. 644, 48080 Bilbao, Spain}
\author{G. Garc\'\i a-Calder\'on}
\altaffiliation{Permanent address: Instituto de F\'{\i}sica, Universidad Nacional Aut\'onoma de M\'exico, Apartado Postal {20 364}, 01000
M\'exico, D.F., M\'exico} \email{gaston@fisica.unam.mx} \affiliation{Departamento de Qu\'\i mica-F\'\i sica, Universidad del Pa\'\i s Vasco,
Apdo. 644, 48080 Bilbao, Spain}
\date{\today}

\begin{abstract}
We describe the influence of continuous measurement in a decaying system and the role of the distance from the detector to the initial location
of the system. The detector is modeled first by a step absorbing potential. For a close and strong detector, the decay rate of the system is
reduced; weaker detectors do not modify the exponential decay rate but suppress the long-time deviations above a coupling threshold.
Nevertheless, these perturbing effects of measurement disappear by increasing the distance between the initial state and the detector, as well
as by improving the efficiency of the detector.
\end{abstract}
 \maketitle

%%%%%%%%%%%%%%%%%%%%%%%%%%%%%%%%%%%%%%%%%%%%%%%%%%%%%%%%%%%%%%%
\section{Introduction}
The decay of unstable quantum states is an ubiquitous process in virtually all fields of physics and energy ranges, from particle and nuclear
physics to condensed matter, or atomic and molecular science. The exponential decay, by far the most common type, is surrounded by deviations at
short and long times \cite{Khalfin57,Fonda72}. The short-time deviations have been much discussed, in particular in connection with the Zeno
effect \cite{Zeno1,Zeno2,Zeno3} and the anti-Zeno effect \cite{Schieve89,Kofman96,Kofman00,Facchi01}. Experimental observations of short
\cite{Raizen97,Raizen01} and long \cite{R06} time deviations are very recent.
%has not been reported so far
%\cite{Norman88,Norman95}.
A difficulty for the experimental verification of long-time deviations has been the weakness of the decaying signal \cite{Gaston01}, but also
the measurement itself may be responsible, because of the suppression of the initial state reconstruction \cite{Fonda72,Muga05}.

It was soon recognized that the measurement could perturb in an important way the dynamics, not only at long times but also at short times, and
that even the rate of exponential decay in the intermediate regime could be affected \cite{Fonda72}. A related and interesting issue is the {\it
quantum Zeno paradox}: repeated instantaneous measurements over a decaying system freeze the decay as the period tends to zero, if the
projection postulate is applied. The same conclusions hold in some limits when ``quantum measurement theory'', which incorporates the measuring
apparatus or part of it in the theoretical model \cite{VonNewman32,Landau31,Ozawa03,Koshino05}, is used for describing generalized
(non-instantaneous and non-ideal) measurements. Several works have analyzed the conditions for the existence of Zeno and anti-Zeno effects in
unstable system  from the point of view of the spectral properties of the response function of the detector \cite{Shaji04,Elattari00,Koshino05}.
For a detailed discussion about quantum Zeno and anti-Zeno effects in a generalized sense (i.e., the slowdown or speedup of the decay for
generalized measurements including continuous ones), see a complete review by K. Koshino and A. Shimizu \cite{Koshino05}.
%In this paper the ``Zeno effect''
%is interpreted in this generalized sense.

A still controversial and rather crucial question is: how is the decay affected by the distance between detector and system in indirect
measurements? \cite{CP94,CP96,Ryff97,HW97,Hotta04,Makris04,Wallentowitz05,Koshino05,Ozawa06}. Home and Whitaker in their conceptual analysis of
the Zeno effect \cite{HW97}, considered that the only really paradoxical point is that the system is predicted to have its decay affected by a
detector at a macroscopic distance. Indeed, a common sense expectation is that separating the detector from the initial location of the system
will soften the perturbing effects of measurement, making the Zeno effect  eventually irrelevant, but the results based on some measurement
models showing this fact have been disputed \cite{CP94,CP96,Wallentowitz05}, and the need for more work has been stated to arrive at more
definite conclusions \cite{Koshino05}.

In the present letter the decaying particle is initially localized within an interaction potential region, and the effect of the ``continuous''
detection is modeled by an imaginary absorbing potential
% $-iV_c\Theta(x-\Delta)$, with $\Delta\ge 1$.
which accounts for the passage from the initial channel to some other channels which are not represented explicitly \cite{ReviewC04}.
%Therefore, as we shall show, the Zeno effect can be observed
%also in non-hermitian, decaying, hamiltonian systems, where not explicit reference %to detectors is made.
A physical system that may be represented in this way is an atom detected by the fluorescence induced by an on-resonance laser beam. The
absorbing potential becomes, in the laser region and in a small lifetime limit, $-i\hbar\Omega^2/2\Gamma$, where $\Omega$ is the Rabi frequency
and $\Gamma$ the inverse of the life time \cite{RDNMH}. In this case, the wave function describes undetected atoms, and every detected atom (by
the  first spontaneous photon detection) ceases to be part of the statistical ensemble associated with the wave function. Moreover, the rate of
norm loss becomes equal to the detection rate \cite{DEHM}. Two helpful approximations for an analytical treatment are a sharply defined beam
edge, and the substitution of the actual beam width by a semi-infinite potential \cite{RDNMH}. The physical validity of this approximation was
studied in \cite{Dambo}, and requires that the penetration length of the undetected atom amplitude be smaller than the laser beam width. This
condition and the production of sharp borders are well within the scope of current ultra-cold atom experiments.

We shall describe the main effects of the absorption (or, as discussed previously, detection), in the decay process of a quantum system. In
particular, we will show that the slowdown of the decay, i.e. the generalized Zeno effect, disappears when the distance to the detector is
increased. We shall also analyze the suppression of the deviations from exponential decay at long times as a function of the strength and
quality of the absorber and the influence of the ``observation distance''.
%
%
%%%%%%%%%%%%%%%%%%%%%%%%%%%%%%%%%%%%%%%%%%%%%%%%%%%%%%%%%%%%%%%%%

\section{Model}
Let us consider a one-dimensional system described by the following (dimensionless) Hamiltonian
\beqa {H}=-\frac{\partial^2}{\partial x^2}+\eta \delta(x-1)-iV_c \Theta(x-\Delta), \qquad x\ge 0, \label{hamil} \eeqa
where $\Theta$ is the Heaviside function, $\Delta$ is the position of the detector edge, $\eta>0$ and $V_c\ge 0$ as corresponds to absorption.
The norm in the absorbing region, for the atom at rest, would decay with a lifetime, or response time of the detector, $1/2V_c$. Because of the
position dependence, however, a very large $V_c$ leads to reflection without detection. Boundary effects can be to a large extent avoided, as we
shall see below, with an adequate shaping of the potential \cite{ReviewC04}.

We assume that the initial state at $t=0$ is the ground state of an infinite well between $x=0$ and $x=1$, namely,
\begin{equation}
\psi(x,0) =\sqrt{2}\sin(\pi x)\,\Theta(x)\Theta(1-x). \label{psi_0}
\end{equation}
Hence, the system  evolves with time according to the non-hermitian Hamiltonian of Eq. (\ref{hamil}),
$ \psi(t)=  e^{-iH t} \psi(0)$.
The survival amplitude, defined as $A(t)=\la \psi(0)|\psi(t)\ra$, may be written in term of the eigenstates forming a biorthogonal basis
\cite{biorthogonal},
\beqa A(t)&=&\sum_{l=1}^{N_{loc}} \mathcal{C}_l\hat{\mathcal{C}}_l\, e^{-iE_lt} \nonumber \\[.3cm]
&+& \int_{0}^{\infty} f(q)\, e^{-i(q^2 -iV_c)t}\; dq\;, \label{evo} \eeqa
where $\mathcal{C}_l=\langle \psi_0|u_l\rangle$, $ \hat{\mathcal{C}}_l=\langle \hat{u}_l|\psi_0\rangle$,
$f(q)=\langle \psi_0|\phi_q \rangle \langle \hat{\phi}_q|\psi_0\rangle$, and
$|u_l\rangle$  and $|\hat{u}_l \rangle $ are, respectively,  right and left localized eigenstates obeying
\beqa {H}|u_l\rangle &=&E_l|u_l \rangle =k_l^2|u_l \rangle, \\[.3cm] \langle \hat{u}_l| {H} &=&E_l \langle \hat{u}_l| =k_l^2\langle \hat{u}_l|,
\\[.3cm] \langle u_l|\hat{u}_j\rangle&=&\delta_{l,j}, \eeqa
$\delta_{l,j}$ being the Kronecker delta and $N_{loc}$ the total number of localized, (Kronecker) normalizable states.
The continuum eigenstates appearing above, $|\phi_q \rangle$ and $|\hat{\phi}_q\rangle$, satisfy
\beqa {H}|\phi_q\rangle &=&E_q|\phi_q \rangle =(q^2-iV_c)|\phi_q \rangle,
\\[.3cm] \langle \hat{\phi}_q| {H} &=&E_q \langle \hat{\phi}_q|=(q^2-iV_c)\langle \hat{\phi}_q|,
\\[.3cm] \langle \phi_q|\hat{\phi}_{q'}\rangle&=&\delta(q-q'). \eeqa
Note that $|\phi_q \rangle$ and its corresponding biorthogonal partner are not usual scattering states because the exterior region is not free
from interaction ($V(x) \ne 0$ when $x\to\infty$). However, the potential is constant there and this enables us to write the solution in the
external region in terms of an ${\cal{S}}$ matrix,
\begin{equation}
\phi_q(x)=\frac{1}{(2\pi)^{1/2}} \left \{
\begin{array}{cc}
C_1\sin\, kx,& 0\leq x\leq 1
\\[.3truecm]
Ae^{ikx}+Be^{-ikx}, & 1\leq x\leq \Delta
\\[.3cm]
e^{-iqx}-{\cal{S}}(q)e^{iqx},& x\geq \Delta
\end{array}
\right. \label{form}
\end{equation}
where $k=(q^2-iV_c)^{1/2}$ is the wavenumber inside, $q$ the wavenumber outside, and $C_1$, $A,B$, and ${\cal{S}}$ are obtained from the
matching conditions at $x=1$ and $x=\Delta$. For scattering-like solutions, $q$ is positive. Note the two branch points of $k$ in the complex
$q$ plane. We shall take the branch cut joining these points. Similarly, the root in $q=(k^2+iV_c)^{1/2}$ is defined with a branch cut joining
the two branch points in the $k$ plane. In contrast to scattering-like states of the continuum, localized states are characterized by a complex
$q$ with positive imaginary part.
%
%%%%%%%%%%%%%%%%%%%%%%%%%%%%%%%%%%%%%%%%%%%%%%%%%%%%%%%%%%%%%%%%%%%%%%

\section{Results}
We shall start analyzing the effects of detection when the detector is placed close to the unstable system, at $\Delta=1$. We have chosen
$\eta=5$ to facilitate the calculations since small values of $\eta$ lead to smaller lifetimes and a badly defined exponential regime whereas,
by contrast, large values of $\eta$ are associated with long lifetimes and well defined (narrow and isolated) resonances, but the numerical
integration of Eq. (\ref{evo}) is much more complicated.

\begin{figure}
\begin{center}
\includegraphics[angle=-90,width=0.778\linewidth]{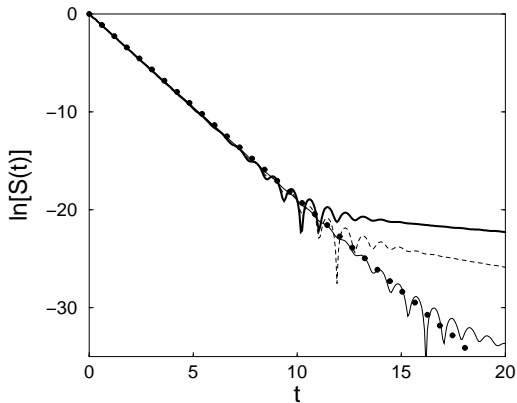}
\end{center}
\caption{\label{fig1}$\ln[S(t)]$ for different absorptive step potentials, see Eq. (\ref{hamil}). $V_c=0$ (thick solid line), $0.1$ (dashed
line), $0.3$ (thin solid line) and $0.5$ (dots). $\Delta=1$ and $\eta=5$.}
\end{figure}

\begin{figure}
\begin{center}
\includegraphics[angle=-90,width=0.778\linewidth]{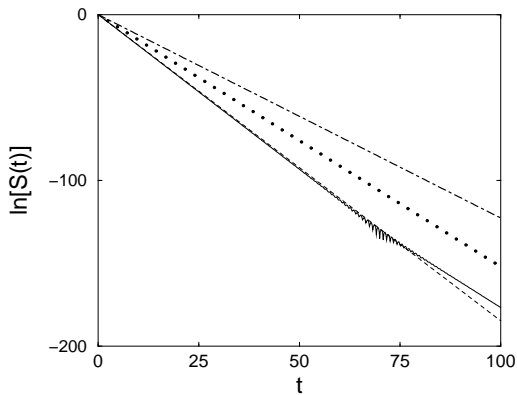}
\end{center}
\caption{\label{fig2}$\ln[S(t)]$ for stronger absorptive potentials (compared to Fig. \ref{fig1}): $V_c=0.75$ (solid line, still with a visible
long-time deviation), $1$ (dashed line), $10$ (dots) and $100$ (dotted-dashed line). $\Delta=1$ and $\eta=5$, as in Fig. \ref{fig1}.}
\end{figure}

In Figs. \ref{fig1} and \ref{fig2}, we have plotted the logarithm of the survival probability, $S(t)=|A(t)|^2$, versus time for different values
of the complex absorbing potential $V_c$. For increasing $V_c$, below a threshold, there is a continuous shift to higher values of the
transition time, $t_{tran}$, which marks the passage from the exponential dominated regime to the final non-exponential decay. Beyond the
threshold value, $V_c^{thre}\approx 0.926$, the decay does not present apparently any deviation from the exponential decay.

\begin{figure}
\begin{center}
\includegraphics[angle=-90,width=0.778\linewidth]{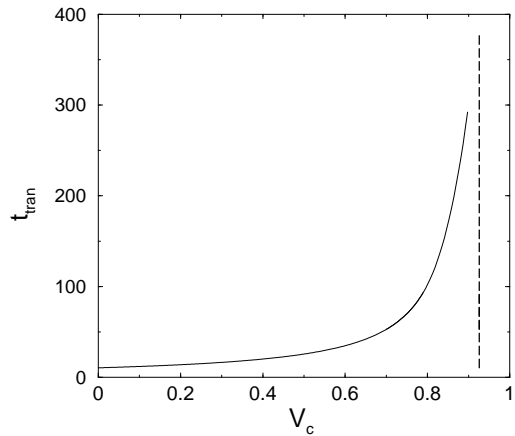}
\end{center}
\caption{\label{fig3}Transition time $t_{tran}$ versus $V_c$ for $\eta=5$ and $\Delta=1$. The critical value $V_c^{thre}$ is marked with a
vertical dashed line.}
\end{figure}

A quantitative approximation to $S(t)$ helps to understand these effects: Let $q_r$ be the resonance with the longest lifetime
and let us assume that it is narrow an isolated. If the rest of the resonances have already decayed, for weak enough absorption, i.e.,
$N_{loc}=0$, the integral of Eq. (\ref{evo}) can be approximated, by using an appropriate contour deformation in the complex  $q$ plane, by the
residue corresponding to the first resonance plus a saddle contribution,
\beqa A(t)&\approx &-2\pi i\,\mbox{Res}\left[f(q)\right]_{q=q_r} e^{-i\mathcal{E}_rt} e^{-\Gamma_rt/2} \nonumber \\[.3truecm]
&&-\frac{\sqrt{\pi i}}{8} {\ddot f}(0)e^{-V_c\,t}\frac{1}{t^{\,3/2}}, \label{apro} \eeqa
where $\mathcal{E}_r$ represents the energy of the decaying particle, $\Gamma_r$ the corresponding decaying width
and \,\,${\ddot f(0)}=[{d^{\,2}f(q)/dq^2}]_{q=0}$.
%%%%%%%%%%%%%%%%%%%%%%%  begin Figure 1  %%%%%%%%%%%%%%%%%%%%%%%%%%%%%%%%%
%%%%%%%%%%%%%%%%%%%%%%%%  end figure 1   %%%%%%%%%%%%%%%%%%%%%%%%%%%%%%%
%
%
%%%%%%%%%%%%%%%%%%%%%%%  begin Figure 2  %%%%%%%%%%%%%%%%%%%%%%%%%%%%%%%%%
%%%%%%%%%%%%%%%%%%%%%%%%  end figure 2   %%%%%%%%%%%%%%%%%%%%%%%%%%%%%%%
%
The second term is responsible for the deviation from the exponential decay in $S(t)$. The novelty with respect to the non-absorption case, $V_c=0$,
is that the deviation is not given by a purely algebraic term: the usual algebraic dependence is multiplied by the exponentially decaying factor
$\exp(-V_ct)$. By increasing $V_c$, the deviation term decays more and more rapidly until, at threshold, \textit{i.e.}, $\Gamma_r=V_c$,  the deviation
decays faster than the residue term. This threshold value corresponds exactly to the passage from a resonance to a localized, normalizable state
with purely exponential decay. While for $V_c < V_c^{thre}$, the dominant term at long times is the saddle contribution (proportional to
$\exp(-V_ct)\,t^{-3/2}\,)$, in the opposite case, $V_c > V_c^{thre}$, the decay is purely exponential, see Fig. \ref{fig2}, and the dominant
contribution comes from the discrete part of the spectrum. This peculiar behavior can be also observed in the divergence of $t_{tran}$ versus
$V_c$ at $V_c^{thre}$, see Fig. \ref{fig3}.
%
%%%%%%%%%%%%%%%%%%%%%%%  begin Figure 3  %%%%%%%%%%%%%%%%%%%%%%%%%%%%%%%%%
%%%%%%%%%%%%%%%%%%%%%%%%  end figure 3   %%%%%%%%%%%%%%%%%%%%%%%%%%%%%%%%%
%
If the absorption is increased further, see Fig. \ref{fig2}, the decay rate decreases, an evidence of a generalized Zeno effect.

\begin{figure}
\begin{center}
\includegraphics[angle=-90,width=0.778\linewidth]{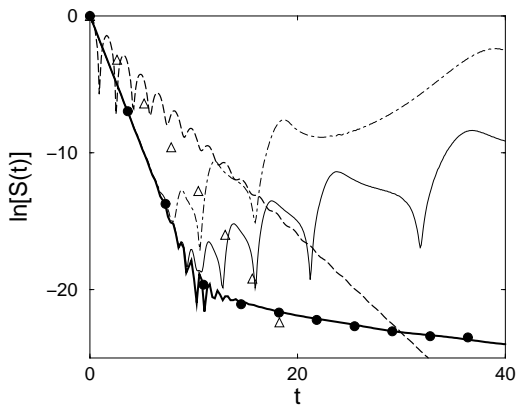}
\end{center}
\caption{\label{fig4}$\ln[S(t)]$ versus time for $\eta=5$, $V_c=100$, and different values of $\Delta$: $1$ (triangles),  $2$ (thick dashed
line),  $100$ (dotted-dashed line) and $200$ (thin solid line). The reference curve for $V_c=0$ (thick solid line) is also shown. The dots
correspond to a better detector placed at $\Delta=100$, see Eq. (\ref{mano}) and related text, with $E_{min}=0.003$ and $L=100$.}
\end{figure}

We shall next examine how the above perturbing effects of measurement are affected by the distance to the absorber. In Fig. \ref{fig4} we show
the dependence of $\ln [S(t)]$ with $\Delta$ for $V_c=100$, compared to the reference case $V_c=0$. For $\Delta=1$, the decay is slowed down
with respect to the exponential decay for $V_c=0$, as in the cases shown in Fig. 2. A small increase of the distance to $\Delta=2$ leads,
perhaps counter-intuitively, to an even slower exponential decay, but only after an oscillatory transient at short times. This is explained by
the approach of a second pole and their mutual interference. By increasing $\Delta$ the poles move closer and closer to each other and a simple
analysis in terms of one or few poles becomes soon impossible. For larger values of $\Delta$ the Zeno effect disappears, namely, the exponential
decay rate is, at least initially, the same as for $V_c=0$, and the transition time $t_{tran}$ separating the exponential decay region and the
long-time deviation increases with $\Delta$, and approaches the transition time for $V_c=0$. Beyond the time region in which the curves with and
without absorption agree, there are some important oscillations in $\ln [S(t)]$,
%
%%%%%%%%%%%%%%%%%%%%%%%  begin Figure 4  %%%%%%%%%%%%%%%%%%%%%%%%%%%%%%%%%
%%%%%%%%%%%%%%%%%%%%%%%%  end figure 4  %%%%%%%%%%%%%%%%%%%%%%%%%%%%%%%%%
%
and a return to the initial state due to the reflectivity at low energies of the sharply edged detector model. To show the importance of the
detector quality we have also used a better absorber, namely, the potential proposed by Manolopoulos \cite{Manolopoulos,ReviewC04},
\beqa V^M(x)=-i{E_{min}}\Theta(x-\Delta) \vartheta\left[\frac{c(x-\Delta)}{L}\right], \label{mano} \eeqa
where $L$ is the absorption width and
\beqa \vartheta(X)=aX-bX^3+\frac{4}{(c-X)^2}-\frac{4}{(c+X)^2}, \nonumber \eeqa
with $a \approx 0.11245$, $b \approx 8.28772\times 10^{-3}$ and $c \approx 2.62206$. For this new model, the survival probability curves (an
example is shown in Fig. \ref{fig4}), fit to the unperturbed curve for a longer time and the back-reaction of the measurement apparatus on the
system is much reduced.

\subsection{Absence of anti-Zeno effect}
\begin{figure}
\begin{center}
\includegraphics[angle=-90,width=0.70\linewidth]{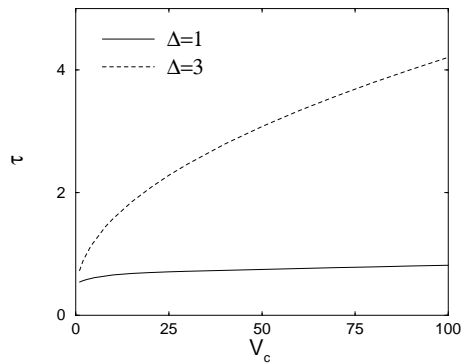}
\end{center}
\caption{\label{fig5} Monotonic increase of the lifetime versus $V_c$ for $\eta=5$ and two detector distances.}
\end{figure}

\begin{figure}
\begin{center}
\includegraphics[width=0.878\linewidth]{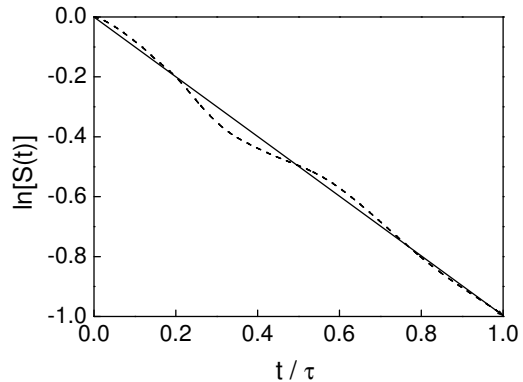}
\end{center}
\caption{\label{fig6} Behaviour of the survival probability versus time in units of the lifetime $\tau$ at short times for $\eta=1$ and $V_c=0$
(dashed-line). The continuous line exhibits, for comparison, pure exponential decay}
\end{figure}

The exponential decay for the values of $V_c$ chosen in Figs. \ref{fig2} and \ref{fig4}  is slower than for $V_c=0$. We have carried out a
more systematic calculation, sweeping continuously over $V_c$: Fig. \ref{fig5} shows the lifetime of the dominant exponential decay versus
$V_c$ and clearly no anti-Zeno effect is observed. This may appear contradictory with the fact that the survival probability for the case
$V_c=0$ decays indeed faster than purely exponential decay in several time spans, see Fig. \ref{fig6}. In more
detail, continuous measurements are usually related to repeated instantaneous measurements of period $\delta t$ by means of Schulman's relation
$\delta t=4/\tau_0$ \cite{Schulman}, where $\tau_0$ is the response time of the apparatus. Since in the time regions of faster decay, the
anti-Zeno effect occurs with repeated instantaneous measurements \cite{Facchi01}, this relation suggests that the continuous measurement should
lead to an anti-Zeno effect for some value of the interaction. That this is not the case shows that Schulman's relation is not directly
applicable in this model. This is because the model Hamiltonian considered by Schulman \cite{Schulman} and ours have different forms and
parameters, in particular there is no $x$, $\Delta$ or $\lambda$ dependency in \cite{Schulman}, where a two channel model is considered, whereas
ours refers to a one channel treatment.

%%%%%%%%%%%%%%%%%%%%%%%  begin Figure 5  %%%%%%%%%%%%%%%%%%%%%%%%%%%%%%%%%
%%%%%%%%%%%%%%%%%%%%%%%%  end figure 5   %%%%%%%%%%%%%%%%%%%%%%%%%%%%%%%
%
%%%%%%%%%%%%%%%%%%%%%%%  begin Figure 6  %%%%%%%%%%%%%%%%%%%%%%%%%%%%%%%%%
%%%%%%%%%%%%%%%%%%%%%%%%  end figure 6   %%%%%%%%%%%%%%%%%%%%%%%%%%%%%%%
%
%%%%%%%%%%%%%%%%%%%%%%%%%%%%%%%%%%%%%%%%%%%%%%%%%%%%%%%%%%%%%%%%%%%%%%%

\section{Concluding remarks}
We have studied the influence of the detector, modeled by a ``step'' negative imaginary potential, in an unstable decaying system. One of the
effects of absorption is the suppression of the deviations from exponential decay at long times, but only above a critical value of the
absorption potential and if it occurs close to the system. The slowdown of the decay rate for strong absorption (Zeno effect) and its dependence
with the distance to the absorber have also been examined: the change in the decay rate due to the continuous measurement is washed out by
increasing the distance to the detector. The perturbing effects of measurement are also reduced by more efficient, reflectionless detectors.
%
%%%%%%%%%%%%%%%%%%%%%%%%%%%%%%%%%%%%%%%%%%%%%%%%%%%%%%%%%%%%%%%%%%%%%%%

\begin{acknowledgments}
This work has been supported by Ministerio de Educaci\'on y Ciencia (BFM2003-01003), and UPV-EHU (00039.310-15968/2004). G. G-C. acknowledges
partial financial support from El Ministerio de Educaci\'on y Ciencia, Spain, under grant No. SAB2004-0010.
\end{acknowledgments}

\end{document}